# Beyond Vibrationally Mediated Electron Transfer: Coherent Phenomena Induced by Ultrafast Charge Separation

*Robert Huber†, Lars Dworak‡, Jacques E. Moser§, Michael Grätzel# and Josef Wachtveitl\*‡*

†Institut für Biomedizinische Optik, Universität zu Lübeck, Peter-Monnik-Weg 4, 23562 Lübeck, Germany

‡Institute of Physical and Theoretical Chemistry, Goethe-University Frankfurt, Max-von-Laue-Str. 7, 60438 Frankfurt/M., Germany

§Photochemical Dynamics Group, Institute of Chemical Sciences and Engineering, and Lausanne Centre for Ultrafast Spectroscopy (LACUS), Ecole Polytechnique Fédérale de Lausanne, CH-1015 Lausanne, Switzerland

#Laboratory for Photonics and Interfaces, Institute of Chemical Sciences and Engineering, Ecole Polytechnique Fédérale de Lausanne, CH-1015 Lausanne, Switzerland

AUTHOR INFORMATION

**Corresponding Author**

\* Institute of Physical and Theoretical Chemistry, Goethe-University Frankfurt, Max von Laue-Str. 7, 60438 Frankfurt/M., Germany, Tel.: +49 (0)69 798 29351, E-Mail: wveitl@theochem.uni-frankfurt.de

ABSTRACT. Wave packet propagation succeeding electron transfer (ET) from alizarin dye molecules into the nanocrystalline $TiO_2$ semiconductor has been studied by ultrafast transient absorption spectroscopy. Due to the ultrafast time scale of the ET reaction of about 6 fs the system shows substantial differences to molecular ET systems. We show that the ET process is not mediated by molecular vibrations and therefore classical ET theories lose their applicability. Here the ET reaction itself prepares a vibrational wave packet and not the electromagnetic excitation by the laser pulse. Furthermore, the generation of phonons during polaron formation in the $TiO_2$ lattice is observed in real time for this system. The presented investigations enable an unambiguous assignment of the involved photoinduced mechanisms and can contribute to a corresponding extension of molecular ET theories to ultrafast ET systems like alizarin/$TiO_2$.

INTRODUCTION

Molecular electron transfer (ET) belongs to the most important and ubiquitously encountered processes in the fields of chemistry and biology. The standard theoretical treatment for molecular ET was developed by Marcus[1,2] providing a correlation of the difference in Gibb's free energy (ΔG) between donor and acceptor, the curvature of the potential energy surfaces (PES) and the reorganization energy (λ). Within the framework of this non-adiabatic theory, ET rates of many systems were predicted quite well. Further quantum mechanical extensions by theories of Hopfield[3] and Jortner[4] were then even able to further extend the scope of theoretical ET models. All of these ET theories are based on the assumption, that the energies of the donor and the acceptor states are matched by energy fluctuations caused by the surrounding solvent acting as thermal bath.

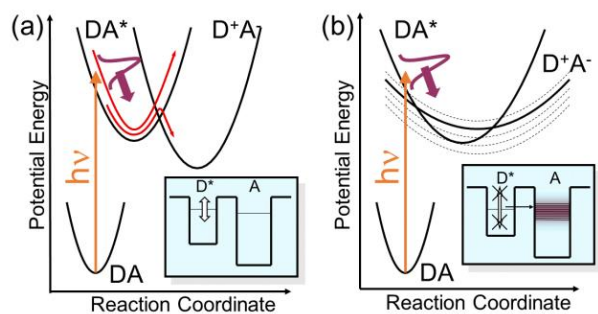

**Figure 1.** Sketch of (a) bimolecular ET and (b) ET at molecule/semiconductor interfaces in the potential energy surface and the electronic quantum dot picture (inset).

In a standard photoinduced ET reaction a bimolecular donor (D) - acceptor (A) system is excited from the ground state DA into the Franck-Condon region of a higher electronic state DA* (Figure 1 (a)). In a classical view the system can subsequently propagate along the PES of the electronic configuration DA*, approximated as one dimensional parabola. Energy conservation allows the electronic transition between the states DA* and $D^+A^-$ only at the intersection of the two PESs. In the picture of electronic quantum dots the propagation of the system on the PES of the state DA* leads to a periodic modulation of the electronic energy level of the donor (Figure 1 (a), inset), corresponding to the potential energy in the PES picture. In the configuration where the energies of donor and acceptor levels match (i.e. exactly at the crossing point of the DA* and the $D^+A^-$ parabolas), ET can occur as a tunneling process with a certain probability, depending on the electronic coupling matrix element between the state DA* and the charge separated state $D^+A^-$. Based on this microscopic view of ET, macroscopic rates can be calculated by assuming a thermal occupation of the energy eigenstates of the DA* potential. Thus, within the Marcus picture, the overall ET process is mediated by molecular vibrations.

For photoinduced ET reactions at molecule-semiconductor interfaces a fundamentally different situation is found resulting in a dramatic change of the energetics and ET kinetics.[5-16] Such reactions play a crucial role in photocatalysis,[17-20] the photographic process[21] and, most importantly, dye/semiconductor systems applied in dye-sensitized solar cells.[22-25] In these systems the dye acts as electron donor which can be excited by a photon usually in the visible spectral range whereas the acceptor level is the energetically broad conduction band of the semiconductor. As a consequence the $D^+A^-$ energy surface splits up into a manifold of acceptor parabolas, so an energy matching mechanism via molecular vibrations is obsolete for the situation depicted in Figure 1 (b). However, there are also dye-semiconductor systems where vibrational motion facilitates the ET. In that case the ET occurs preferentially each time the vibrational wavepacket reaches an intersection of the DA* PES and an acceptor parabola of the $D^+A^-$ manifold leading to a step-like ET process.[9]

Obviously, interface systems differ fundamentally from conventional molecular ET arrangements and the underlying processes have been studied extensively by quantitative computational analysis for different dye/semiconductor systems including alizarin/$TiO_2$.[26-31]

Here, we present investigations on the dye alizarin coupled onto $TiO_2$ colloidal nanocrystals. As this dye/semiconductor system exhibits an ultrafast photoinduced ET with a time constant of about 6 fs,[11] it provides the ideal prerequisites for studies on the coherent coupling of the ET reaction to molecular and semiconductor lattice vibrations, because the charge separation occurs much faster than any significant molecular motion. For the assignment of the observed oscillations, the nonreactive reference system alizarin/$ZrO_2$ is also investigated and the results are compared to those of alizarin/$TiO_2$. We will show, that in the case of alizarin/$TiO_2$ the ET is not mediated by molecular vibrations but here the ET-associated charge separation itself excites coherent molecular oscillations, a process beyond the traditional Marcus theory of ET.

EXPERIMENTAL METHODS

**Sample preparation**. The preparation of the dye/semiconductor systems has been published elsewhere.[8,11] Briefly, $ZrO_2$ and $TiO_2$ colloidal nanoparticles have been obtained from hydrolysis of $TiCl_4$ and $ZrCl_4$. The investigated alizarin/$ZrO_2$ and alizarin/$TiO_2$ samples were prepared from the colloidal nanoparticles and had final alizarin concentrations of 0.5 mM.

**Transient absorption spectroscopy**. The dye/semiconductor systems were investigated as colloidal solutions using a conventional femtosecond pump/probe setup.[8,12,32] The transient absorbance changes after excitation were recorded by super-continuum probe pulses generated in a $CaF_2$ plate which covered a spectral range from 400 to 650 nm. For excitation we used pump pulses from a noncollinear optical parametric amplifier.[33] The excitation wavelength was adjusted to 495 nm according to the spectral position of the alizarin absorption band. The system had a temporal resolution of 20-30 fs dependent on the probe wavelength. A detailed description of the experimental setup can be found in Ref. 11.

RESULTS AND DISCUSSION

The spectral properties of the investigated samples as well as the transient absorption dynamics related to the population and depopulation of electronic states can be found in elsewhere.[8,11] Here, we exclusively focus on the oscillatory patterns observed in time resolved measurements which have not been subject of our earlier studies.

The periodic deformation of a molecule due to a classical oscillatory motion, e.g. triggered by photoexcitation with an ultrashort laser pulse, can cause a periodic spectral shift of the absorption and emission bands of the molecule in the particular electronic state what leads to a frequency modulated dynamic spectrum.[34-36] Consequently the absorbance changes observed in femtosecond experiments at a certain spectral position are then a superposition of the exponential kinetics of electronic population and depopulation dynamics and the overlaid oscillatory kinetics due to the molecular vibrations. Subtracting a fit curve satisfying the exponential characteristics of the transients yields the oscillatory residua, the real time signal of structural molecular vibrations.

The 2D spectra of the residual oscillations obtained for the alizarin/$ZrO_2$ and the alizarin/$TiO_2$ system are shown in Figure 2. The alizarin/$ZrO_2$ 2D spectrum has oscillation patterns in the spectral range between 400-550 nm (Figure 2 (a)). The oscillation observed for the alizarin/$TiO_2$ system exhibits larger amplitudes and a significantly different wavelength-dependence with stronger contributions at longer wavelengths (Figure 2 (b)).

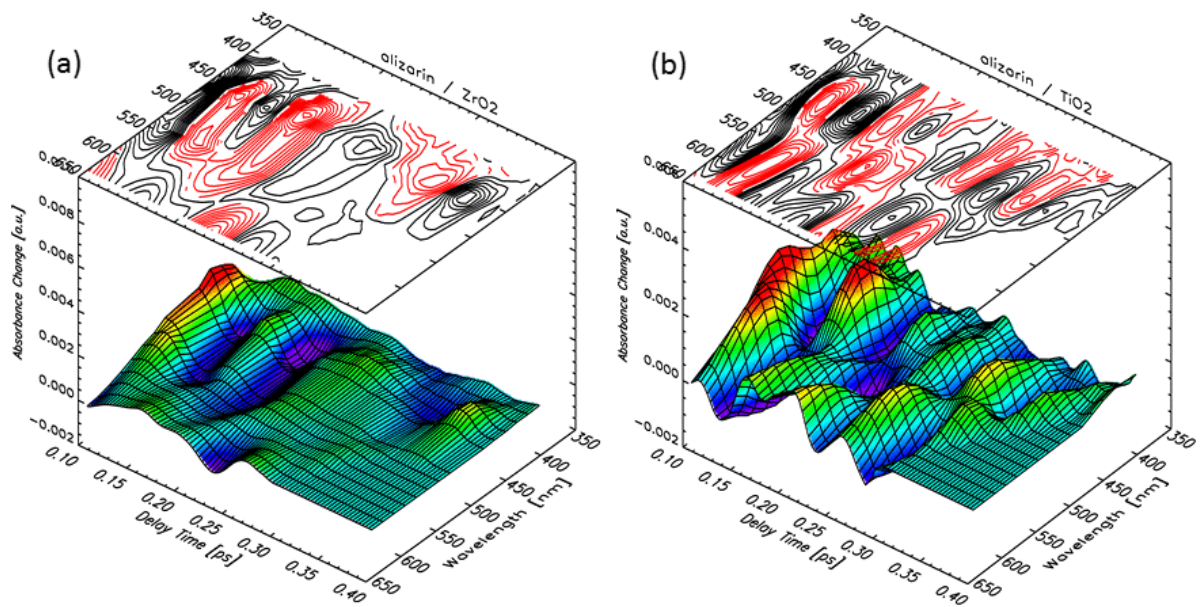

**Figure 2.** 2D spectra of the residual oscillations for (a) the alizarin/$ZrO_2$ and (b) the alizarin/$TiO_2$ system.

The extracted oscillatory signals at different wavelengths are evaluated by a global Fourier transformation, resulting in the Fourier transformed (FT) spectra for alizarin/$ZrO_2$ and alizarin/$TiO_2$ shown in Figure 3.

On the basis of identical experimental conditions in both cases, $ZrO_2$ and $TiO_2$, all deviations between the two presented spectra should reflect differences in the reaction mechanism. As previously described, in the case of $ZrO_2$ no electron injection into the conduction band can be observed, at most a partial population of energetically low lying surface states.[8] Thus, the comparison of the Fourier spectra of the two systems allows a distinction between different kinds of vibrational normal modes with respect to their excitation mechanism. Two different types of molecular vibrations can be discerned, corresponding either to oscillations directly excited by the applied laser pulse or representing the response of the dye molecule and the semiconductor nanocrystal upon the ET reaction.

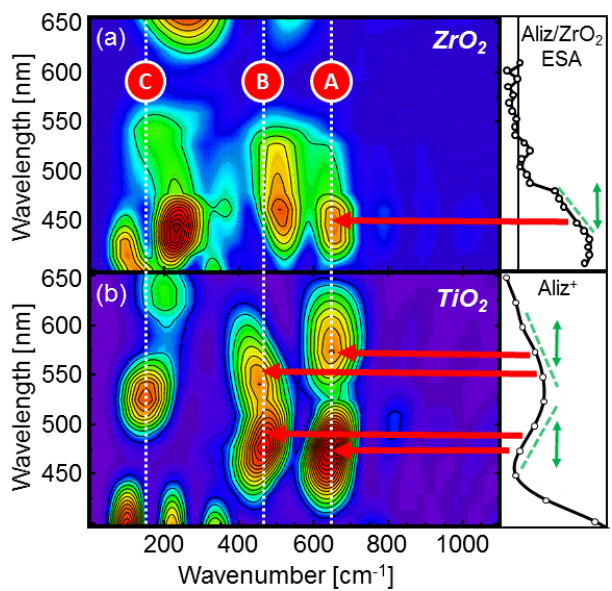

**Figure 3.** Fourier transformed spectra of (a) the alizarin/ZrO$_2$ system and (b) the alizarin/TiO$_2$ system. The three marked frequencies ("A", "B', "C") assign molecular eigenmodes with their amplitude maxima (vertical lines). (a), right: Excited state absorption band obtained from transient absorption measurements on alizarin/ZrO$_2$ (see ref. 12); (b), right: absorption spectrum of alizarin cation (see ref. 8). The arrows indicate which peak in the 2D plots correspond to the respective slopes in the absorption spectra.

*Type I: Oscillations excited by the laser pulse*

Oscillatory contributions with an explicit intensity in the FT spectra at ~650 cm$^{-1}$ (Figure 3, mode labeled "A") for both, the alizarin/ZrO$_2$ and the alizarin/TiO$_2$ system are observed. In the case of ZrO$_2$ the correlated FT spectrum in Figure 3 (a) shows a maximum in the spectral region where the slope in the alizarin excited state absorption (ESA) band (Figure 3 (a), right) is large. This observation is indicative of a frequency modulated transient absorption spectrum due to a periodic spectral shift of the alizarin ESA band. In the difference spectrum, the strongest modulation is observed at spectral positions where the slope in the absorption of the oscillating species is large. Therefore, the modes detected for alizarin/ZrO$_2$ indeed reflect wave packet propagations in the alizarin excited state. The alizarin/TiO$_2$ system undergoes an ET reaction much faster than the time scale of any molecular motion ($\tau_{ET}$ = 6 fs, corresponding to a frequency of 5000 cm$^{-1}$). So in the corresponding FT spectrum (Figure 3 (b)) the amplitude of the prepared mode at ~650 cm$^{-1}$ exhibits maxima at the spectral positions where the slope in the absorption spectrum of the alizarin cation (Figure 3 (b), right) is large. Considering that the determined frequencies are almost identical for both investigated systems, the same generation mechanism can be expected. Hence, for the alizarin/TiO$_2$ system the mode labeled "A" is a vibrational molecular normal mode excited by the laser pulse surviving the ET process undisturbed. This mode preserves full vibrational coherence during the ET reaction which gives experimental evidence that molecular ET can occur keeping a fixed phase relation of the nuclear wave function. The comparable frequencies of 647 cm$^{-1}$ for alizarin/TiO$_2$ and 659 cm$^{-1}$ for alizarin/ZrO$_2$ imply similar geometries of the alizarin excited state and the alizarin cation.

In Figure 3 (a) additional modes are found for alizarin/ZrO$_2$ at ~220 cm$^{-1}$ and ~500 cm$^{-1}$ which are not observed for the alizarin/TiO$_2$ system. The amplitude maxima of these modes coincides with the alizarin ESA at ~450 nm indicative of wave packet propagations in the alizarin excited state.

Ground state oscillations, excited by an impulsive stimulated Raman scattering (ISRS) process, were not found in our system. These modes would be easy to identify, since they should have an identical appearance in the FFT-spectra for the TiO$_2$ as well as for the ZrO$_2$ system with a spectral signature related to the alizarin ground state absorption.

*Type II: Oscillations excited by the ET reaction – coherent response of the system*

Besides vibrational modes of "type I", reflecting the evolution of a quantum mechanical wave packet generated by the laser pulse as already observed for other adsorbates,[37] the FT-spectrum of

alizarin/TiO$_2$ in Figure 3 (b) reveals a fundamentally different type of vibrational mode: At a frequency of 460 cm$^{-1}$ in the alizarin/TiO$_2$ system two prominent peaks occur (Figure 3, labeled as "B"), which are absent in the nonreactive alizarin/ZrO$_2$ system. It can therefore be concluded, that the mechanism for preparing this kind of oscillatory wave packet is not the photoexcitation with the ultrashort laser pulse and the related S-S* transition of the alizarin, but the ET reaction itself. Hence, these measurements allow the direct observation of a vibrational wave packet, generated by a chemical (ET) reaction and not by the impact of the electromagnetic field of a pulsed laser. This mechanism is only possible for systems reacting much faster than the vibrational dephasing of the relevant modes.

In contrast to the observed "type I" oscillations the wave packet does not only survive the ET reaction, but the ET-related charge transfer prepares a coherent superposition of vibrational modes by projecting the population from the S*-PES to the S$^+$/TiO$_2$-PES of the charge separated state. As the S$^+$/TiO$_2$ PES is expected to lie asymmetrically above the S$_0$/TiO$_2$ ground state (cf. Figure 1), the wave packet starts to propagate towards the new energy minimum. In a picture of electron density this scenario corresponds to a situation, where the excitation from the S$_0$ to the S* state changes the electron density in a way, that the equilibrium molecular geometry is not changed in this special normal coordinate. However, after the ultrafast ET reaction the electronic configuration of the alizarin cation is characterized by a new equilibrium configuration (with respect to the normal coordinate for the 460 cm$^{-1}$ mode). Consequently, the alizarin cation will start to oscillate around the new equilibrium geometry. This is corroborated by the wavelength-dependence of this mode illustrated in Figure 3 (b). For alizarin/TiO$_2$ the mode at 460 cm$^{-1}$ thus directly reflects the coherent response of the molecular structure to the preceding ET process. Earlier quantum dynamical calculations on the dye/semiconductor system coumarin 343/TiO$_2$ indicated that the ultrafast ET can indeed prepare wave packets on the PES of the acceptor states, which then start to oscillate.[38]

The pronounced peak at 145 cm$^{-1}$ in the TiO$_2$ system (Figure 3, labeled as "C") correlates almost perfectly with the most dominant Raman-band found in anatase TiO$_2$.[39,40] The origin of a transiently modulated transmittance of the actually transparent medium TiO$_2$ in a spectrally and temporally resolved measurement can be explained by a time dependent modulation of the refractive index of TiO$_2$ due to lattice vibrations. This effect of electronic coherence and impulsive stimulated Raman scattering for spectrally broad probe pulses was extensively described by Kovalenko et al..[41]

Additional experiments on coumarin 343 coupled to the surface of TiO$_2$ and ZrO$_2$ nanocrystals have been conducted. In analogous measurements on coumarin 343/TiO$_2$ exactly the same mode at 145 cm$^{-1}$ is observed (Figure 4 (a)) despite the different nature of the dye whereas neither in the system alizarin/ZrO$_2$ nor in coumarin/ZrO$_2$ (Figure 4 (b)) any oscillation at this frequency could be found. The oscillation at 145 cm$^{-1}$ thus represents ET-induced lattice vibrations of the TiO$_2$ crystal. This observation is in line with earlier studies which revealed charge separation induced lattice vibrations for PbSe quantum dot/TiO$_2$ heterostructures,[42] CdSe quantum dots[43] and quantum dot/methylviologen complexes.[35] Based on ultraviolet photoemission spectroscopy and two-photon photoemission Gundlach et al. reported a vibrational excited ionized perylene dye along with a broad energy distribution of electrons injected to the TiO$_2$.[44,45] In contrast to these investigations, wave packets in real time at distinct probe wavelengths are measured in our study. The results allows to unambiguously trace the source of the observed vibrational modes.

In Figure 4 (b) a mode with moderate amplitude at 175 cm$^{-1}$ is observed for alizarin/ZrO$_2$ what is in good agreement with a dominant mode at 180 cm$^{-1}$ measured for monoclinic ZrO$_2$.[46] Since electron transfer from photoexcited alizarin to the ZrO$_2$ conduction band is not possible the observed mode is probably related to the ZrO$_2$ nanoparticle surface. It is known that electron transfer to surface trap states can occur on the fs time-scale.[8]

While in an energetic picture the electron is transferred from the molecular energy level into the conduction band of the TiO$_2$ nanocrystal, in coordinate space this transition corresponds to a real displacement of electronic probability density from the molecule to the nanocrystal with a subsequent distortion of the TiO$_2$ crystal lattice due to the altered electrostatic field. In other words, this distortion can be described by a more or less pronounced self-trapping mechanism of the electron due to the formation of a polaron. The ultrafast time scale of the injection of the additional charge and the correlated formation of a polaron leads to a coherent excitation of lattice vibrations, i.e. of phonons, which can then be detected by transient spectroscopy. So the mode at 145 cm$^{-1}$ is a pure semiconductor property and represents the coherent response of the electron acceptor on the changed charge distribution after the ET. This mode of the TiO$_2$ lattice is the equivalent to the molecular oscillation of the alizarin cation at 460 cm$^{-1}$ regarding their excitation mechanism.

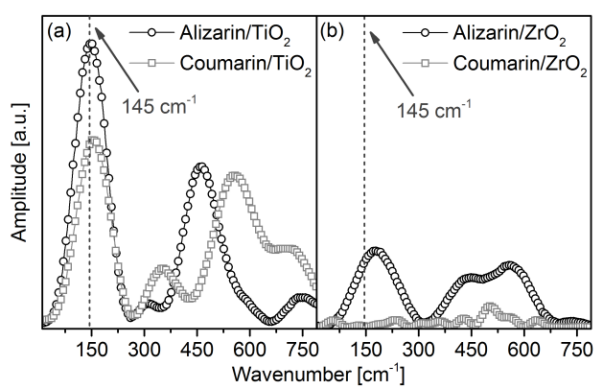

**Figure 4.** Fourier transformed data of (a) alizarin/TiO$_2$ and coumarin/TiO$_2$ samples as well as (b) alizarin/ZrO$_2$ and coumarin/ZrO$_2$ samples at a probe wavelength of 526 nm. The black line indicates the mode at 145 cm$^{-1}$.

To rule out that the observed oscillations are directly caused by optical excitation of the $TiO_2$ though ISRS, a solution of colloidal $TiO_2$ nanocrystals without coupled dye was measured. It showed negligible oscillatory signals implying that the oscillation was not triggered by the laser pulse itself.

An excitation mechanism caused by the distribution of electron density in the S* state can be ruled out, as the excited state of the adsorbed alizarin persists only for about 6 fs, a time much shorter than the observed 145 $cm^{-1}$ molecular oscillation period. The observed vibration can hardly be excited by the altered electron configuration upon excitation because the total momentum will be small due to the short lifetime of the excited state. This assumption is substantiated by the fact, that in the two nonreactive $ZrO_2$ systems no common mode is found.

Phonon emission during the cooling process of the injected, hot electron as reason for the mode excited at 145 $cm^{-1}$ can also be ruled out. The cooling process does not lead to a coherent excitation of lattice vibrations, as cooling times lie on a time scale of several hundreds of femtoseconds, much longer than the observed oscillation period. Here, like in most cases, cooling is a non-coherent process.

CONCLUSIONS

In conclusion, propagating wave packets are identified, both on the donor and the acceptor side of an ultrafast ET system, which are prepared directly by the ET process and not by the initial photoexcitation. In contrast to usual molecular ET, ultrafast reactions as presented here are not mediated by molecular oscillations but, on the contrary, they themselves generate a coherent superposition of vibrational eigenmodes. The findings underline the need to further expand ET-theories for a valid description of reactions on the sub-10-fs time scale found at molecule/solid state interfaces.

AUTHOR INFORMATION

**Notes**

The authors declare no competing financial interest.

ACKNOWLEDGMENT

We thank W. Zinth for helpful discussions. This work was supported by the German Research Foundation (DFG) (Hu 1006/6-1, WA 1850/6-1) and European Union projects FDML-Raman (FP7 ERC StG, contract no. 259158) and ENCOMOLE-2i (Horizon 2020, ERC CoG no. 646669). J.E.M. and M.G. are grateful to the Swiss National Science Foundation (FNRS) for financial support.

REFERENCES

(1) Marcus, R. A. Chemical and Electrochemical Electron-Transfer Theory. *Annu. Rev. Phys. Chem.* **1964**, *15*, 155-196.

(2) Marcus, R.A. Electron-Transfer Reactions in Chemistry - Theory and Experiment (Nobel Lecture). *Angew. Chem., Int. Ed. Engl.* **1993**, *32,* 1111-1121.

(3) Hopfield, J. *Electrical Phenomena at the Biological Membrane Level*. Elsevier Sci. Publ: Amsterdam, 1977.

(4) Jortner, J. Temperature-Dependent Activation-Energy for Electron-Transfer between Biological Molecules. *J. Chem. Phys.* **1976**, *64*, 4860-4867.

(5) Moser, J.-E.; Grätzel, M. Photosensitized Electron Injection in Colloidal Semiconductors. *J. Am. Chem. Soc.* **1984**, *106*, 6557-6564.

(6)  Vinodgopal, K.; Hua, X.; Dahlgren, R. L.; Lappin, A. G.; Patterson, L. K.; Kamat, P. V. Photochemistry of Ru(bpy)$_2$(dcbpy)$^{2+}$ on Al$_2$O$_3$ and TiO$_2$ Surfaces. An Insight into the Mechanism of Photosensitization. *Phys. Chem.* **1995**, *99*, 10883-10889.

(7)  Burfeindt, B.; Hannappel, T.; Storck, W.; Willig, F. Measurement of Temperature-independent Femtosecond Interfacial Electron Transfer from an Anchored Molecular Electron Donor to a Semiconductor as Acceptor. *J. Phys. Chem.* **1996**, *100*, 16463-16465.

(8)  Huber, R.; Spörlein, S.; Moser, J.-E.; Grätzel, M.; Wachtveitl, J. The Role of Surface States in the Ultrafast Photoinduced Electron Transfer from Sensitizing Dye Molecules to Semiconductor Colloids. *J. Phys. Chem. B* **2000**, *104*, 8995-9003.

(9)  Zimmermann, C.; Willig, F.; Ramakrishna, S.; Burfeindt, B.; Pettinger, B.; Eichberger, R.; Storck, W. Experimental Fingerprints of Vibrational Wave-Packet Motion During Ultrafast Heterogeneous Electron Transfer. *J. Phys. Chem. B* **2001**, *105*, 9245-9253.

(10) Benkö, G.; Kallioinen, J.; Korppi-Tommola, J. E. I.; Yartsev, A. P.; Sundström, V. Photoinduced Ultrafast Dye-to-Semiconductor Electron Injection from Nonthermalized and Thermalized Donor States. *J. Am. Chem. Soc.* **2002**, *9*, 489-493.

(11) Huber, R.; Moser, J.-E.; Grätzel, M.; Wachtveitl, J. Real-time Observation of Photoinduced Adiabatic Electron Transfer in Strongly Coupled Dye/Semiconductor Colloidal Systems with a 6 fs Time Constant. *J. Phys. Chem. B* **2002**, *106*, 6494-6499.

(12) Huber, R.; Moser, J.-E.; Grätzel, M.; Wachtveitl, J. Observation of Photoinduced Electron Transfer in Dye/Semiconductor Colloidal Systems with Different Coupling Strengths. *Chem. Phys.* **2002**, *285*, 39-45.

(13) Schnadt, J.; Brühwiler, P. A.; Patthey, L; O'Shea, J. N.; Södergren, S.; Odelius, M.; Ahuja, R.; Karis, O.; Bässler, M.; Persson, P.; et al. Experimental Evidence for sub-3-fs Charge Transfer from an Aromatic Adsorbate to a Semiconductor. *Nature* **2002**, *418*, 620-623.

(14) Dworak, L.; Matylitsky, V. V.; Wachtveitl, J. Ultrafast Photoinduced Processes in Alizarin-Sensitized Metal Oxide Mesoporous Films. *ChemPhysChem* **2009**, *10*, 384-391.

(15) Nomoto, T.; Fujio, K.; Sasahara, A.; Okajima, H.; Koide, N.; Katayama, H.; Onishi, H. Phonon Mode of TiO$_2$ Coupled with the Electron Transfer from N3 Dye. *J. Chem. Phys.* **2013**, *138*, 224704-224709.

(16) Rittmann-Frank, M. H.; Milne, C. J.; Rittmann, J.; Reinhard, M.; Penfold, T. J.; Chergui, M. Mapping of the Photoinduced Electron Traps in TiO$_2$ by Picosecond X-ray Absorption Spectroscopy. *Angew. Chem. Int. Ed.* **2014**, *53*, 5858-5862

(17) Asahi, R.; Morikawa, T.; Ohwaki, T.; Aoki, K.; Taga, Y. Visible-Light Photocatalysis in Nitrogen-Doped Titanium Oxides. *Science* **2001**, *293*, 269-271.

(18) Jiang, D. L.; Zhao, H. J.; Zhang, S. Q.; John, R. Kinetic Study of Photocatalytic Oxidation of Adsorbed Carboxylic Acids at TiO$_2$ Porous Films by Photoelectrolysis. *J. Catal.* **2004**, *223*, 212-220.

(19) Zhao, W.; Ma, W. H.; Chen, C. C.; Zhao, J. C.; Shuai, Z. G. Efficient Degradation of Toxic Organic Pollutants with Ni$_2$O$_3$/TiO$_{2-x}$B$_x$ under Visible Irradiation. *J. Am. Chem. Soc.* **2004**, *126*, 4782-4783.

(20) Banerjee, S.; Pillai, S. C.; Falaras, P.; O'Shea, K. E.; Byrne, J. A.; Dionysiou, D. D. New Insights into the Mechanism of Visible Light Photocatalysis. New Insights into the Mechanism of Visible Light Photocatalysis. *J. Phys. Chem. Lett.* **2014**, *5*, 2543−2554.

(21) Liu, D.; Hug, G. L.; Kamat, P. V. Photochemistry on Surfaces - Intermolecular Energy and Electron-Transfer Processes between Excited Ru(bpy)$_3^{2+}$ and H-Aggregates of Cresyl Violet on SiO$_2$ and SnO$_2$ Colloids. *J. Phys. Chem.* **1995**, *99*, 16768-16775.

(22) Oregan, B.; Grätzel, M. A Low-Cost, High-Efficiency Solar-Cell Based on Dye-Sensitized Colloidal TiO$_2$ Films. *Nature* **1991**, *353*, 737-740.

(23) Bach, U.; Lupo, D.; Comte, P.; Moser J.-E.; Weissortel, F.; Salbeck, J.; Spreitzer, H.; Grätzel, M. Solid-State Dye-Sensitized Mesoporous TiO$_2$ Solar Cells with high Photon-to-Electron Conversion Efficiencies. *Nature* **1998**, *395*, 583-585.

(24) Grätzel, M. Photoelectrochemical Cells. *Nature* **2001**, *414*, 338-344.

(25) Yella, A.; Lee, H.-W.; Tsao, H. N.; Yi, C.; Chandiran, A. K.; Nazeeruddin, M. K.; Wei-Guang Diau, E.; Yeh, C.-Y.; Zakeeruddin, S. M.; Grätzel, M. Porphyrin-Sensitized Solar Cells with Cobalt (II/III)–Based Redox Electrolyte Exceed 12 Percent Efficiency. *Science* **2011**, *334*, 629-634.

(26) Ramakrishna, S.; Willig, F.; May, V. Photoinduced Ultrafast Electron Injection from a Surface Attached Molecule: Control of Electronic and Vibronic Distributions via Vibrational Wave Packets. *Phys. Rev. B* **2000**, *62*, R16330-R16333.

(27) Rego, L. G. C.; Batista, V. S. Quantum Dynamics Simulations of Interfacial Electron Transfer in Sensitized TiO$_2$ Semiconductors. *J. Am. Chem. Soc*. **2003**, *125*, 7989-7997.

(28) Stier, W.; Duncan, W. R.; Prezhdo, O. V. Thermally Assisted Sub-10 fs Electron Transfer in Dye-Sensitized Nanocrystalline TiO$_2$ Solar Cells. *Adv. Mater*. **2004**, *16*, 240-244.

(29) Duncan, W. R.; Stier, W.; Prezhdo, O. V. Ab Initio Nonadiabatic Molecular Dynamics of the Ultrafast Electron Injection across the Alizarin-TiO$_2$ Interface. *J. Am. Chem. Soc*. **2005**, *127*, 7941-7951.

(30) Li, J.; Kondov, I.; Wang, H.; Thoss, M. Theoretical Study of Photoinduced Electron-Transfer Processes in the Dye-Semiconductor System Alizarin-TiO$_2$. *J. Phys. Chem. C* **2010**, *114*, 18481–18493.

(31) Akimov, A. V.; Neukirch, A. J.; Prezhdo, O. V. Theoretical Insights into Photoinduced Charge Transfer and Catalysis at Oxide Interfaces. *Chem. Rev*. **2013**, *113*, 4496−4565.

(32) Seel, M.; Wildermuth, E.; Zinth, W. A Multichannel Detection System for Application in Ultra-Fast Spectroscopy. *Meas. Sci. Technol*. **1997**, *8*, 449-452.

(33) Huber, R.; Satzger, H.; Zinth, W.; Wachtveitl, J. Noncollinear Optical Parametric Amplifiers with Output Parameters Improved by the Application of a White Light Continuum Generated in $CaF_2$. *Opt. Commun*. **2001**, *194*, 443-448.

(34) Sagar, D. M.; Cooney, R. R.; Sewall, S. L.; Dias, E. A.; Barsan, M. M.; Butler, I. S.; Kambhampati, P. Size Dependent, State-Resolved Studies of Exciton-Phonon Couplings in strongly Confined Semiconductor Quantum Dots. *Phys. Rev. B* **2008**, *77*, 235321.

(35) Dworak, L.; Matylitsky, V. V.; Braun, M.; Wachtveitl, J. Coherent Longitudinal Optical Ground-State Phonon in CdSe Quantum Dots Triggered by Ultrafast Charge Migration. *Phys. Rev. Lett.* **2011**, *107*, 247401.

(36) Schweighöfer, F.; Dworak, L.; Braun, M.; Zastrow, M.; Wahl, J.; Burghardt, I.; Rück-Braun, K.; Wachtveitl, J. Vibrational Coherence Transfer in an Electronically Decoupled Molecular Dyad. *Sci. Rep*. **2015**, *5*, 9368.

(37) Ramakrishna, S.; Willig, F.; May, V. Photoinduced Ultrafast Electron Injection from a Surface Attached Molecule: Control of Electronic and Vibronic Distributions via Vibrational Wave Packets. *Phys. Rev. B* **2000**, *62*, R16330-R16333.

(38) Thoss, M.; Kondov, I.; Wang, H. Correlated Electron-Nuclear Dynamics in Ultrafast Photoinduced Electron-Transfer Reactions at Dye-Semiconductor Interfaces. *Phys. Rev. B* **2007**, *76*, 153313-153316.

(39) Ohsaka, T.; Izumi, F.; Fujiki, Y. Raman-Spectrum of Anatase, $TiO_2$. *J. Raman Spectrosc*. **1978**, *7*, 321-324.

(40) Likodimos, V.; Stergiopoulos, T.; Falaras P.; Harikisun, R.; Desilvestro, J.; Tulloch, G. Prolonged Light and Thermal Stress Effects on Industrial Dye-Sensitized Solar Cells: A Micro-Raman Investigation on the Long-Term Stability of Aged Cells. *J. Phys. Chem. C* **2009**, *113*, 9412–9422.

(41) Kovalenko, S. A.; Dobryakov, A. L.; Ruthmann, J.; Ernsting, N. P.: Femtosecond Spectroscopy of Condensed Phases with Chirped Supercontinuum Probing. *Phys. Rev. A* **1999**, *59*, 2369-2384

(42) Tisdale, W. A.; Williams, K. J.; Timp, B. A.; Norris, D. J.; Aydil, E. S.; Zhu, X.-Y. Hot-Electron Transfer from Semiconductor Nanocrystals. *Science* **2010**, *328*, 1543-1547.

(43) Tyagi, P.; Cooney, R. R.; Sewall, S. L.; Sagar, D. M.; Saari, J. I.; Kambhampati, P. Controlling Piezoelectric Response in Semiconductor Quantum Dots via Impulsive Charge Localization. *Nano Lett*. **2010**, *10*, 3062-3067.

(44) Gundlach, L.; Letzig, T.; Willig, F. Test of Theoretical Models for Ultrafast Heterogeneous Electron Transfer with Femtosecond Two-photon Photoemission Data. *J. Chem. Sci.* **2009**, *121*, 561-574.

(45) Gundlach, L.; Burfeindt, B.; Mahrt, J.; Willig, F. Dynamics of Ultrafast Photoinduced Heterogeneous Electron Transfer, Implications for Recent Solar Energy Conversion Scenarios. *Chem. Phys. Lett*. **2012**, *545*, 35-39.

(46) Kim, B.-K.; Hamaguchi, H. Mode Assignments of the Raman Spectrum of Monoclinic Zirconia by Isotopic Exchange Technique. *Phys. Stat. Sol.* **1997**, *203*, 557-563.

**TOC GRAPHICS**

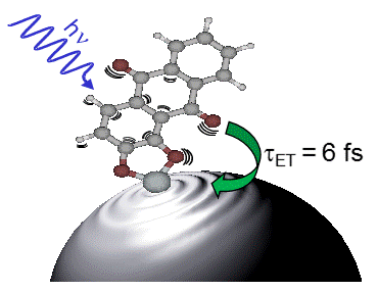